\documentclass[12pt]{article}
\usepackage{amsmath,amssymb,epsf}

\def\nn{\nonumber}

\let\bm=\bibitem

\newcommand{\be}{\begin{equation}}
\newcommand{\ee}{\end{equation}}
\def\ba{\begin{array}}
\def\ea{\end{array}}
\def\ft#1#2{{\textstyle{\frac{\scriptstyle #1}{\scriptstyle #2}}}}
\def\fft#1#2{\frac{#1}{#2}}

\def\sst#1{{\scriptscriptstyle #1}}

\def\td{\tilde}

\def\dalemb#1#2{{\vbox{\hrule height .#2pt
        \hbox{\vrule width.#2pt height#1pt \kern#1pt
                \vrule width.#2pt}
        \hrule height.#2pt}}}

\newcommand{\bea}{\begin{eqnarray}}
\newcommand{\eea}{\end{eqnarray}}

\def\0{{\sst{(0)}}}
\def\1{{\sst{(1)}}}
\def\2{{\sst{(2)}}}
\def\3{{\sst{(3)}}}
\def\4{{\sst{(4)}}}
\def\5{{\sst{(5)}}}
\def\6{{\sst{(6)}}}
\def\7{{\sst{(7)}}}
\def\8{{\sst{(8)}}}

\def\R{\rlap{\rm I}\mkern3mu{\rm R}}

\def\R{\rlap{\rm I}\mkern3mu{\rm R}}

\def\R{{{\mathbb R}}}

\begin{document}
\begin{flushright}
{\bf hep-th/0605296}\\
May\  2006
\end{flushright}

\vspace{30pt}

\begin{center}

{\Large {\bf Enhancing the Jet Quenching Parameter\\ from Marginal
Deformations}}

\vspace{20pt}

J.F. V\'azquez-Poritz

\vspace{20pt}

{\it Department of Physics\\
University of Cincinnati\\ Cincinnati OH 45221-0011, USA}

\vspace*{0.8cm} \centerline{\tt jporitz@physics.uc.edu}

\vspace{40pt}

\underline{ABSTRACT}
\end{center}

A number of recent papers have applied the AdS/CFT correspondence
to a strong-coupling calculation of the medium-induced radiative
parton energy loss in nucleus-nucleus collisions at RHIC. The
predicted value of the ``jet quenching parameter'' ${\hat q}$,
however, is rather small compared to the experimental results. For
hot ${\cal N}=4$ supersymmetric Yang-Mills theory, certain
marginal deformations can have the effect of enhancing ${\hat q}$.
This result is highly sensitive to the location of the fundamental
string's endpoints in the internal space.

\newpage

\section{Introduction}

Collisions of nuclei at RHIC may shed light on the properties of
hot and dense QCD matter. In particular, the plasmic medium
modifies the fragmentation of partons that are produced with high
transverse momentum \cite{baier,kovner,gyulassy,jacobs}. This
medium-dependent effect is described by the ``jet quenching
parameter'' ${\hat q}$. The so-called dipole approximation which
is frequently used in jet quenching calculations is given by
\be \langle W^A ({\cal C})\rangle \approx {\rm exp} [-\ft14 {\hat
q} L^- L^2]\,, \label{q} \ee
where $\langle W^A ({\cal C})\rangle$ is the expectation value of
the light-like Wilson loop in the adjoint representation whose
contour ${\cal C}$ is a rectangle with large extension $L^-$ in
the $x^-$ direction and small extension $L$ in a transverse
direction \cite{zakharov}. It has been proposed that (\ref{q}) can
be taken to be a nonperturbative definition of ${\hat q}$
\cite{liu}.

Furthermore, since the quark-gluon plasma produced at RHIC is
believed to be strongly-coupled, it has been argued that the
AdS/CFT correspondence \cite{malda} is a suitable framework with
which to calculate the energy loss of quarks
\cite{herzog1,liu,solana,buchel,gubser,herzog2,caceres}. On the
supergravity side, it is more convenient to calculate the thermal
expectation value of a Wilson loop in the fundamental
representation $\langle W^F ({\cal C})\rangle$. In the planar
limit, $\langle W^F ({\cal C})\rangle$ and $\langle W^A ({\cal
C})\rangle$ are related by
\be \langle W^A ({\cal C})\rangle =\langle W^F ({\cal
C})\rangle^2\,. \label{AF} \ee
According to the AdS/CFT correspondence,
$\langle W^F ({\cal C})\rangle$ is given by
\be \langle W^F ({\cal C})\rangle = {\rm exp} [-S({\cal C})]\,,
\label{W} \ee
where $S$ is the regularized action of the extremal surface of a
fundamental string whose large-distance boundary is the contour
${\cal C}$ in Minkowski spacetime $\R^{3,1}$
\cite{wilson1,wilson2,wilson3,wilson4,wilson5}. In principle, the
$\langle W^A ({\cal C})\rangle$ found with (\ref{AF}) and
(\ref{W}) via the AdS/CFT correspondence can then be equated with
its proposed nonperturbative definition in (\ref{q}) in order to
give a theoretical prediction for ${\hat q}$ which can then be
compared the RHIC measurements.

Although we do not yet have a supergravity description of
strongly-coupled QCD, we are able to study various gauge theories
at large $N_c$ from the supergravity point of view. The
most-studied example is the ${\cal N}=4$ $SU(N_c)$ supersymmetric
Yang-Mills theory which in the planar limit at large 't Hooft
coupling is described by type IIB supergravity on AdS$_5\times
S^5$ \cite{malda}. In fact, it is for this case that the
calculation of the jet quenching parameter was done in \cite{liu}.
However, the predicted value is rather small compared to the
experimental results. This need not be worrisome, since this
calculation was done for a superconformal field theory as a
preliminary demonstration.

In fact, it has been found that the jet quenching parameter is
gauge theory specific and not universal \cite{buchel}. Therefore,
it is of interest to consider the predictions for the value of
${\hat q}$ in theories which are more realistic, such as those
that are non-conformal at zero temperature and have less
supersymmetry. This was done in \cite{buchel}, for example, in the
case of a strongly-coupled non-conformal gauge theory plasma.
However, the resulting ${\hat q}$ was found to decrease, rather
than increase, as one goes away from the conformal gauge theory.

One way in which the supersymmetry of the theory in the zero
temperature limit can be reduced to ${\cal N}=1$ is to replace the
5-sphere with other Sasaki-Einstein spaces, such as $T^{1,1}$
\cite{T11} or one of the countably infinite $Y^{p,q}$ \cite{Ypq}
or $L^{p,q,r}$ spaces \cite{Lpqr}. One could also replace the
5-sphere by an Einstein space which does not support Killing
spinors, such as the $T^{pq}$ spaces, in order to have a
nonsupersymmetric theory at zero temperature \cite{Tpq}. However,
since the Wilson loop of a purely radial string configuration is
independent of the internal space, this will not affect the
calculation for ${\hat q}$.

We will consider an alternative way by which the supersymmetry can
be reduced down to ${\cal N}=1$, which is to have the theory
undergo marginal deformations. Since the theory has an isometry
group which includes $U(1) \times U(1)$, one to use U-duality to
find the gravity dual of the deformed theory. The deformation on
the gravity side can be matched to an exactly marginal operator in
the field theory, providing a holographic test of the methods of
Leigh and Strassler \cite{LS}. In particular, the deformed
superpotential is given by
\be W=Tr(e^{i\pi\beta} \Phi_1\Phi_2\Phi_3- e^{-i\pi\beta}
\Phi_1\Phi_3\Phi_2)\,, \ee
where $\beta=\gamma-\tau_s \sigma$, $\gamma$ and $\sigma$ are real
deformation parameters and $\tau_s$ is a complex structure
parameter related to the gauge coupling and theta parameter of the
dual gauge theory. The resulting geometry of the gravity dual is a
warped product of AdS$_5$ with the deformed internal space, where
the warp factor depends on the internal directions. As shown in
Figure 1, when temperature is added to the theory, these
deformations can ultimately have the effect of enhancing the ``jet
quenching parameter'' ${\hat q}$.

\begin{figure}[ht]
   \epsfxsize=4.0in \centerline{\epsffile{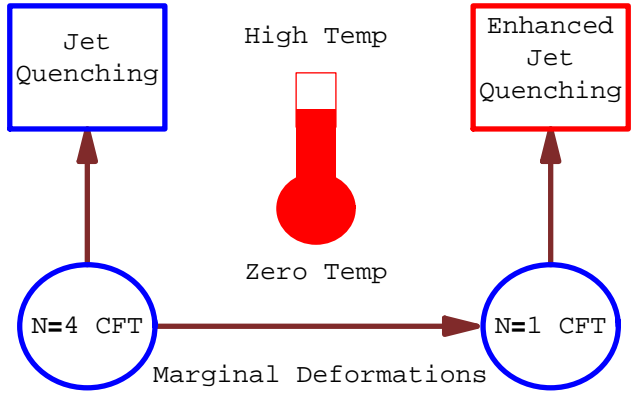}}
   \caption[FIG. \arabic{figure}.]{Marginal deformations of
   conformal field theories can change the properties of the
   theories at finite temperature. In particular, $\sigma$ deformations
   can lead to an enhancement of the "jet quenching parameter" $\hat{q}$.}
   \label{jet1}
\end{figure}

This is actually one of the less dramatic effects that marginal
deformations can have on a theory as it is moved away from its
conformal limit. For example, $\sigma$ deformations of the Coulomb
branch of ${\cal N}=4$ super Yang-Mills theory can induce a
transition from Coulombic attraction between quarks and
anti-quarks to that of linear confinement. This was studied from
the point of view of type IIB supergravity in
\cite{ahn1,ahn2,ahn4}. Similar phenomena were discussed on the
gauge theory side in \cite{dorey1,dorey2,dorey3}.

This paper is organized as follows. In section 2, we will review
the gravity dual of marginal deformations of ${\cal N}=4$
supersymmetric Yang-Mills theory which reduce the supersymmetry
down to ${\cal N}=1$. We will show how $\sigma$ deformations can
enhance the ${\hat q}$ parameter, depending on the location of the
fundamental string endpoints in the internal space. In section 3,
we will generalize this to a more general class of marginal
deformations which do not preserve any supersymmetry. We discuss
various issues and further directions in section 4.

\section{From ${\cal N}=1$ supersymmetric deformations}

Figure 2 shows the steps involved in a solution-generating
technique which can be employed to find the type IIB supergravity
duals of marginally-deformed field theories. The procedure can be
outlined as follows. First, T-dualize along one of the $U(1)$
directions to type IIA theory. Lifting the solution to eleven
dimensions provides a third direction which is associated with a
$U(1)$ symmetry. One can now apply an $SL(3,R)$ transformation
along these $U(1)^3$ directions. Dimensionally reducing and
T-dualizing along shifted directions yields a new type IIB
solution. Of course, instead of lifting to eleven dimensions, one
could apply an $SL(2,R)$ transformation to the type IIA solution.
However, this corresponds to $\gamma$ deformations with $\sigma=0$
which, as we will see, do not effect the Wilson loop calculation
for purely radial strings.

\begin{figure}[ht]
   \epsfxsize=4.0in \centerline{\epsffile{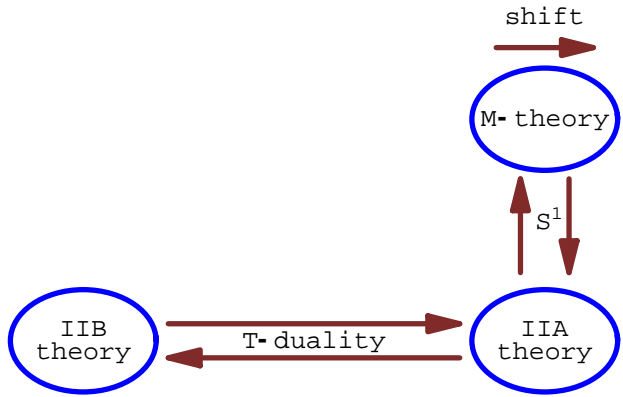}}
   \caption[FIG. \arabic{figure}.]{Generating new solutions via U-dualities.}
   \label{jet2}
\end{figure}

This procedure can be applied to any solution that has an isometry
group which contains $U(1)\times U(1)$. If in addition to this
symmetry there is a $U(1)$ R-symmetry, then the deformed solution
preserves ${\cal N}=1$ supersymmetry. This has been applied to
finding the type IIB supergravity background corresponding to
marginal deformations of ${\cal N}=4$ super Yang-Mills theory
\cite{LM}, as well as the marginal deoformations of the
superconformal field theories associated with the $Y^{p,q}$ spaces
\cite{LM} and the $L^{p,q,r}$ spaces \cite{ahn3}.

One can also apply this solution-generating technique to the
gravity duals of nonsupersymmetric field theories. Here we will
consider the near-horizon of a nonextremal D3-brane. This
corresponds to a Yang-Mills theory at finite temperature, which is
an ${\cal N}=4$ superconformal field theory in the zero
temperature limit \cite{malda}. The metric can be expressed in
light-cone coordinates\footnote{We are using light-cone
coordinates because this will lead to a light-like string
configuration, which ensures that the corresponding quarks are
ultrarelativistic.} as
\be ds_{10}^2=ds_5^2+R^2 d\Omega_5^2\,, \ee
where
\be ds_5^2 = \fft{r^2}{R^2}\Big( -(1+f) dx^+ dx^-+ dx_2^2+dx_3^2 +
\ft12 (1-f) [(dx^+)^2+(dx^-)^2]\Big) +\fft{R^2}{r^2 f}\, dr^2\,.
\ee
where $f=1-\ft{r_0^4}{r^4}$ and $R^4=\alpha^{\prime 2}
\alpha_{{\rm SYM}} N_c$. The metric on the unit 5-sphere can be
written as
\be d\Omega_5^2=\sum_{i=1}^3 (d\mu_i^2+\mu_i^2\,
d\phi_i^2)\,,\qquad\qquad {\rm with}\,\, \sum_i \mu_i^2=1\,. \ee
We can parameterize the 5-sphere metric as
\be \mu_1=\cos\alpha\,,\qquad
\mu_2=\sin\alpha\,\cos\theta\,,\qquad \mu_3=\sin\alpha\,
\cos\theta\,. \ee
The temperature of the dual field theory is equal to the Hawking
temperature of the black hole, which is $T=\fft{r_0}{\pi R^2}$.

The $\beta$ deformation of nonextremal D3-branes in the
near-horizon region has the string-frame metric
\be ds_{str}^2 = \sqrt{H} (ds_5^2+R^2 d{\td\Omega}_5^2)\,,
\label{form} \ee
where
\be d{\td\Omega}_5^2= \sum_i (d\mu_i^2+G\mu_i^2 d\phi_i^2)+
|\hat\beta|^2 \mu_1^2 \mu_2^2 \mu_3^2 (\sum_i d\phi_i)^2\,. \ee
The functions $G$ and $H$ are given by
\bea G^{-1} &=& 1+|{\hat\beta}|^2\, s_{\alpha}^2
(c_{\alpha}^2+s_{\alpha}^2 s_{\theta}^2 c_{\theta}^2)\,,\nn\\
H &=& 1+{\hat\sigma}^2 \, s_{\alpha}^2 (c_{\alpha}^2+s_{\alpha}^2
s_{\theta}^2 c_{\theta}^2)\,, \eea
where $\hat\gamma\equiv \gamma R^2$, $\hat\sigma\equiv \sigma R^2$
and $\hat\beta\equiv \beta R^2$.

If we begin with $N$ D3-branes, then the deformed solution
includes $\gamma N$ D5-branes and $\sigma N$ NS5-branes wrapped on
a two-torus. Charge quantization implies that $\gamma$ and
$\sigma$ must take on rational values. There are also flux and
scalar fields turned on by the deformations but these are not
important for our purposes. What is important is the conformal
factor $H$, which renders the solution implicitly higher
dimensional. Note that $H$ depends on $\sigma$ but not $\gamma$.
Also, note that $H\ge 1$ which, as we will see shortly, is why the
value of the jet quenching parameter ${\hat q}$ is enhanced rather
than diminished.

The classical supergravity description is valid as long as the
curvature is small relative to the string scale, the two-torus
corresponding to the $U(1)^2$ global symmetry is larger than the
string scale, and the metric does not degenerate at arbitrary
points. These conditions can be met for \cite{LM}
\be R>>1\,,\qquad \hat\gamma<<R\,,\qquad \hat\sigma<<R\,.
\label{sugraconditions} \ee

The action for the Euclideanized string worldsheet is given by
\be S=\fft{1}{2\pi\alpha^{\prime}} \int d\tau\, d\sigma\,
\sqrt{{\rm det}\, G_{MN}\,\partial_{\alpha} X^M\, \partial_{\beta}
X^N}\,. \ee
We will set $\tau=x^-$ and $\sigma=x_2$, so that the string lies
at constant $x_3$ and $x^+$. Also, we are interested in a purely
radial string which is lies on a single point in the internal
space. Therefore, the effective metric is
\be ds_{{\rm eff}}^2=\sqrt{H}\Big[ \fft{r^2}{R^2}\, d\sigma^2 +
\fft{r^2}{2R^2} (1-f) d\tau^2+\fft{R^2}{r^2 f}\, dr^2\Big]\,. \ee
In order for such a string configuration to solve the equations of
motion, it is required that
$\partial_{\alpha}H=\partial_{\theta}H=0$. This only occurs at
several particular points in the internal space. The solutions for
which $H\ge 1$ are: \vspace{.5cm}
\begin{center}
\begin{tabular}{|c|c|c|}
\hline $\sin^2\alpha$ & $\sin^2\theta$ & $H$ \\
\hline\hline $1$ & $\fft12$ & $1+\fft14 \hat\sigma^2$ \\
$\fft12$ & $0$, $1$ & \\
\hline\hline $\fft23$ & $\fft12$ & $1+\fft13 \hat\sigma^2$ \\
\hline
\end{tabular}
\end{center}
\vspace{.5cm}

We are interested in the latter cases for which $H>1$, since only
then will the thermal expectation value for purely radial string
configurations differ from that on the undeformed background. Then
\be S=\sqrt{H}\, \fft{r_0^2 L^-}{\sqrt{2} \pi \alpha^{\prime} R^2}
\int_0^{\ft{L}{2}} d\sigma \sqrt{1+\fft{(\partial_{\sigma}r)^2 R^4
}{f r^4}}\,. \ee
Since $H$ is a constant, the Wilson loop calculation proceeds as
in \cite{liu}. For completeness, we will show some of the details.
The radial equation of motion is
\be (\partial_{\sigma}r)^2=c^2 \fft{r^4 f}{R^4}\,, \label{eom} \ee
where $c$ is an integration constant. For the nontrivial solution
with nonvanishing $c$, the turning point occurs at the black hole
horizon where $f=0$, for all values of $L$. A rough picture of
this is given by Figure 3. This is to be expected, since ${\hat
q}$ describes the thermal medium and not the ultraviolet physics
\cite{liu}. The previous condition $R>>1$ ensures that the horizon
is far enough from the black hole singularity so that the
curvature is small and classical gravity can be trusted there.

\begin{figure}[ht]
   \epsfxsize=4.0in \centerline{\epsffile{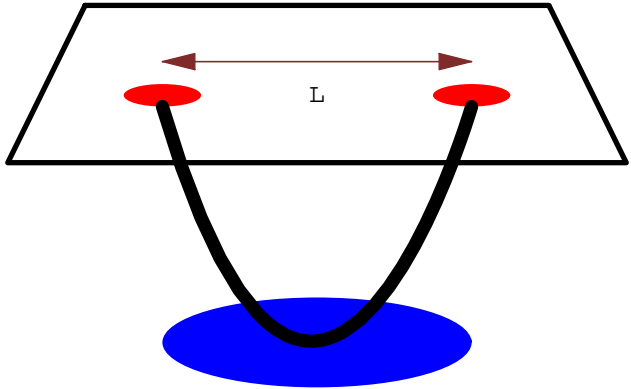}}
   \caption[FIG. \arabic{figure}.]{The thermal expectation value
   of a Wilson loop can be calculated on the supergravity side by
   considering a fundamental string in the background of an AdS
   black hole. The string endpoints lie on a probe brane at large
   distance and the string's turning point is on the black hole
   horizon. However, this picture is slightly misleading because
   the directions along the probe brane are actually orthogonal to
   those along the horizon.}
   \label{jet3}
\end{figure}

Integrating (\ref{eom}) yields $L=2\sqrt{\pi} \Gamma
(\ft54)/(\Gamma (\ft34) R^2cr_0)$ and
\be S=\sqrt{H}\, \fft{\pi \sqrt{\alpha_{{\rm SYM}} N_c} L^- L
T^2}{2\sqrt{2}} \sqrt{1+\fft{4 \Gamma^2 (\ft54)}{\pi \Gamma^2
(\ft34) L^2 T^2}}-\sqrt{H}\, \fft{\sqrt{\pi} \Gamma
(\ft54)}{\sqrt{2} \Gamma (\ft34)} \sqrt{\alpha_{{\rm SYM}}} L^- T
\,. \ee
where we have subtracted the ``self-energy'' of the quark and
antiquark. We can now use (\ref{q}), (\ref{AF}) and (\ref{W}) to
obtain an expression for the jet quenching parameter ${\hat q}$
from the above string action in the limit $LT<<1$. The ${\hat q}$
for the marginally deformed theory can be written in terms of that
for the undeformed theory:
\be {\hat q}_{{\rm deformed}} = \sqrt{H}\, {\hat q}\,, \ee
where ${\hat q}$ was found in \cite{liu} to be
\be {\hat q}=\fft{\pi^{3/2} \Gamma (\ft34)}{\sqrt{2} \Gamma
(\ft54)} \sqrt{\alpha_{{\rm SYM}} N_c}\, T^3\,. \ee
Therefore, the value of the jet quenching parameter is increased
by a factor of $\sqrt{H}$ by the $\sigma$ deformations. The
$\gamma$ deformations do not have any effect in this regard.

One can also consider the $\beta$ deformations of
finite-temperature theories which reduce, in the zero temperature
limit, to the theories associated with the $Y^{p,q}$ spaces and
the $L^{p,q,r}$ spaces. The corresponding gravity dual has the
string-frame metric of the form (\ref{form}) with
$d{\td\Omega}_5^2$ replaced by the deformed $L^{p,q,r}$ metric. As
before, for a purely radial string configuration it is the warp
factor $H$ which is important, rather than the exact form of the
deformed internal metric. Consider the theory associated with
$T^{1,1}$ as an example. After the $\beta$ deformations, the
resulting $H$ is given by
\be H=1+\hat\sigma^2\,\Big(
\fft{1}{54}(\cos^2\theta_2\,\sin^2\theta_1
+\cos^2\theta_1\,\sin^2\theta_2)+\fft{1}{36}\sin^2\theta_1\,\sin^2\theta_2\Big)\,,
\ee
where $\theta_1$ and $\theta_2$ are two coordinates in $T^{1,1}$
\cite{LM}. For a purely radial string configuration, we must have
$\partial_{\theta_1}H=\partial_{\theta_2}H=0$. The solutions for
which $H\ge 1$ are:
\vspace{.5cm}
\begin{center}
\begin{tabular}{|c|c|c|}
\hline $\sin^2\phi_1$ & $\sin^2\phi_2$ & $H$ \\
\hline\hline $1$ & $0$ & $1+\fft{2}{27} \hat\sigma^2$ \\
$0$ & $1$ & \\
\hline\hline $1$ & $1$ & $1+\fft19 \hat\sigma^2$ \\
\hline
\end{tabular}
\end{center}
\vspace{.5cm}

\section{From nonsupersymmetric deformations}

In the previous section, we considered marginally deformed
theories which were supersymmetric in their zero temperature
limits. We found that, for the $\sigma$ deformations, purely
radial string configurations can only exist at particular points
in the internal space. Since there are just a handful of such
points for which there is an enhancement of the ${\hat q}$
parameter, it is important to see if more general backgrounds can
yield more examples for which this enhancement occurs.

We will now consider dropping the condition that our deformed
backgrounds preserve supersymmetry. We will focus on the case in
which the undeformed theory is hot ${\cal N}=4$ super Yang-Mills.
In particular, since the 5-sphere of the initial geometry has
three directions which associated with a $U(1)$ isometry, the
steps described in Figure 2 can be applied three consecutive times
to three different pairs of $U(1)$ directions. The resulting
6-parameter deformed metric has the form (\ref{form}) with
\cite{frolov1,frolov2}
\be d{\td\Omega}_5^2= \sum_i (d\mu_i^2+G\mu_i^2 d\phi_i^2) +G
\mu_1^2 \mu_2^2 \mu_3^2 |\sum_i \beta_i R^2 d\phi_i|^2\,, \ee
where
\bea G^{-1} &=& 1+|\hat\beta_1|^2\mu_2^2\mu_3^2+
|\hat\beta_2|^2\mu_1^2\mu_3^2
+|\hat\beta_3|^2\mu_1^2\mu_2^2\,,\nn\\
H &=& 1+\hat\sigma_1^2\mu_2^2\mu_3^2+ \hat\sigma_2^2\mu_1^2\mu_3^2
+\hat\sigma_3^2\mu_1^2\mu_2^2\,. \eea
Various tests of the AdS/CFT correspondence have been successfully
carried out for this background \cite{frolov1,frolov2}.

The conditions for a purely radial string, namely
$\partial_{\alpha}H=\partial_{\theta}H=0$, imply that
\bea 0 &=& [\hat\sigma_1^2 s_{\alpha}^2 s_{2\theta}^2+2c_{2\alpha}
(\hat\sigma_2^2 s_{\theta}^2+
\hat\sigma_3^2 c_{\theta}^2)]s_{2\alpha}\,,\nn\\
0 &=& [ \hat\sigma_1^2 s_{\alpha}^2 c_{2\theta}+
(\hat\sigma_2^2-\hat\sigma_3^2)c_{\alpha}^2]s_{\alpha}^2
s_{2\theta}\,. \eea
It is straightforward but not very illuminating to solve for the
most general solutions. Here, we present some simple solutions for
which $H>1$, along with the conditions imposed on $\sigma_i$:
\vspace{.5cm}
\begin{center}
\begin{tabular}{|c|c|c|c|}
\hline conditions & $\sin^2\alpha$ & $\sin^2\theta$ & $H$ \\
\hline\hline none & $1$ & $\fft12$ &
$1+\fft14 \hat\sigma_1^2$ \\
\hline none & $\fft12$ & $1$ & $1+\fft14 \hat\sigma_2^2$ \\
$\hat\sigma_1=0$, $\hat\sigma_2=\hat\sigma_3$ & $\fft12$ & not specified & \\
\hline none & $\fft12$ & $0$ &
$1+\fft14 \hat\sigma_3^2$ \\
\hline\hline $\hat\sigma_2=\hat\sigma_3$ &
$\fft{2\hat\sigma_2^2}{4\hat\sigma_2^2-\hat\sigma_1^2}$ & $\fft12$
&
$1+\fft{\hat\sigma_2^4}{4\hat\sigma_2^2-\hat\sigma_1^2}$ \\
\hline $\hat\sigma_1=\hat\sigma_2$ &
$\fft{3\hat\sigma_1^2-\hat\sigma_3^2}{4\hat\sigma_1^2-\hat\sigma_3^2}$
& $ \fft{2\hat\sigma_1^2-\hat\sigma_3^2}{3\hat\sigma_1^2-
\hat\sigma_3^2}$ &
$1+\fft{\hat\sigma_1^4}{4\hat\sigma_1^2-\hat\sigma_3^2}$ \\
\hline $\hat\sigma_1=\hat\sigma_3$ &
$\fft{3\hat\sigma_1^2-\hat\sigma_2^2}{4\hat\sigma_1^2-\hat\sigma_2^2}$
& $\fft{\hat\sigma_1^2}{3\hat\sigma_1^2-\hat\sigma_2^2}$ &
$1+\fft{\hat\sigma_1^4}{4\hat\sigma_1^2-\hat\sigma_2^2}$ \\
\hline
\end{tabular}
\end{center}
\vspace{.5cm}

The reality conditions on $\alpha$ and $\theta$ put further
constraints on the $\hat\sigma_i$ which, in turn, ensure that $H>
1$. As can be seen in the above table, dropping the supersymmetry
condition leads to an increase in the number of string endpoints
in the internal space which lead to an enhancement of the jet
quenching parameter ${\hat q}$. For the case of equal
$\hat\sigma_i$, ${\cal N}=1$ supersymmetry is preserved and the
above solutions degenerate to those of the first example that was
considered in the previous section.

\section{Discussion}

We have considered particular marginal deformations of ${\cal
N}=4$ supersymmetric Yang-Mills theory which, at the level of
supergravity, decrease the supersymmetry down to ${\cal N}=1$ or
${\cal N}=0$. When temperature is turned on, we have demonstrated
that the $\sigma$ deformations can have the effect of enhancing
the jet quenching parameter ${\hat q}$ as follows:
\be {\hat q}\rightarrow \sqrt{1+c^2\,\hat\sigma^2}\,{\hat q}\,,
\ee
where $c$ is a constant which depends on the location of the
fundamental string's endpoints in the internal space. In
particular, $c$ is nonzero only when the endpoints are at
particular points in the internal space. This demonstrates the
that value of the jet quenching parameter ${\hat q}$ can be highly
sensitive to certain deformations of the theory. Probing more
realistic theories with less degrees of freedom might lead to
greater predictive power.

For instance, one can generalize the results of \cite{liu} to
theories which exhibit a confining phase \cite{PS,KS,MN} or
contain fundamental matter \cite{flavor1,flavor2}. A step in the
former direction was done in \cite{buchel}. However, the jet
quenching parameter was found to decrease as one goes from the
conformal gauge theory to a confining gauge theory, thereby
increasing the discrepancy between the predicted value and the
experimental measurements at RHIC. One might wonder if $\sigma$
deformations could help solve this problem. Unfortunately, such
deformations would not lead to a well-defined supergravity
solution, due to the presence of the RR-flux on the 3-sphere
\cite{LM}. On the other hand, the background used in \cite{buchel}
is perturbative in the deformation away from the conformal gauge
theory \cite{temp2,temp3}. In fact, there is no clear reason to
expect that the ${\hat q}$ parameter monotonically increases as
one goes from the conformal phase to QCD. Recently, a numerical
numerical solution was found which might include a nonextremal
generalization of the Klebanov-Strassler solution \cite{leo}.
Since this solution works for all temperatures, it would be
interesting to see if the jet quenching parameter increases as one
gets closer to QCD.

We have only considered purely radial string configurations. One
could also consider strings whose endpoints lie at two different
locations within the internal space. Then the corresponding quarks
will have different scalar charges. One might naively expect that
the string action, and therefore the parameter ${\hat q}$,
increases if the string endpoints move away from each other within
the internal space. This is because there will be additional
positive terms in the action of the form
$(\partial_{\sigma}\phi)^2$, where $\phi$ is an internal
coordinate and $\sigma$ is the spatial coordinate of the string
worldsheet. However, at least for the simplest scenario of an
(undeformed) AdS black hole background,this is more than
compensated for by a decrease in the $(\partial_{\sigma}r)^2$
term. Therefore, moving the string endpoints away from each other
serves to decrease the value of ${\hat q}$. The result might be
different in more complicated backgrounds, such as with the
addition of marginal deformations.

The marginally deformed solutions discussed in this paper may
provide new backgrounds on which to investigate transport
phenomena such as diffusion and sound propagation. It would be
interesting to see if the universal ratio between the shear
viscosity and the entropy density
\cite{universal1,universal2,universal3} is obeyed. It would also
be interesting to calculate the speed of sound in this background,
as was done for the gravity duals of other nonconformal field
theories \cite{nonconformal1,nonconformal2,nonconformal3}.
Unfortunately, it might be rather difficult to apply the
techniques of holographic renormalization to a background whose
asymptotic geometry has a warp factor which depends on the
internal directions.

\vspace{.7cm}

\centerline{\bf Acknowledgments}

We would like to thank Philip Argyres and Leopoldo Pando Zayas for
useful discussions, as well as Alex Buchel and Hong Liu for
helpful correspondence. This research is supported by DOE grant
DOE-FG02-84ER-40153.

\end{document}